\newcommand\nodata{{...} }
\newcommand\eg{{\it e.g.} }

\newcommand\etal{et~al.}

\newcommand\kms{\ifmmode {\rm\,km\,s^{-1}}\else${\rm\,km\,s^{-1}}$\fi}

\font\smallaipsfont = cmsy6 scaled\magstep1
\newcommand\aips {{\smallaipsfont AIPS}}
\def\spose#1{\hbox to 0pt{#1\hss}}
\newcommand\simlt{\mathrel{\spose{\lower 3pt\hbox{$\mathchar"218$}}
     \raise 2.0pt\hbox{$\mathchar"13C$}}}
\newcommand\simgt{\mathrel{\spose{\lower 3pt\hbox{$\mathchar"218$}}
     \raise 2.0pt\hbox{$\mathchar"13E$}}}
\newcommand\aap{{\em A\&A}}
\newcommand\aaps{{\em A\&AS}}
\newcommand\aj{{\em AJ}}
\newcommand\apj{{\em ApJ}}
\newcommand\apjl{{\em ApJ}}

\newcommand\jaa{{\em J. Astrophys. Astr.}}

\newcommand\mnras{{\em MNRAS}}

\def\procspie{Proc.~SPIE}%

\documentclass[usenatbib]{mn2e}

\usepackage{psfig}
\usepackage{rotating}
\usepackage{amssymb}

\title[Distant radio galaxies from SUMSS and NVSS]{A search for
distant radio galaxies from SUMSS and NVSS: I. Sample definition,
radio and $K-$band imaging\thanks{Based on observations obtained with
the Australia Telescope Compact Array, the Anglo-Australian Telescope,
and the European Southern Observatory, La Silla, Chile (Program
70.A-0514)}.}
\author[C. De Breuck \etal]{
\parbox[t]{\textwidth}{
Carlos De Breuck$^1$\thanks{Current address: European Southern Observatory, Karl Schwarzschild Stra\ss e 2, D-85748 Garching, Germany}, Richard W.\
Hunstead$^2$, Elaine M.\ Sadler$^2$, Brigitte Rocca--Volmerange$^1$,
Ilana Klamer$^2$}
\vspace*{6pt} \\ 
$^1$ Institut d'Astrophysique de Paris, 98bis Boulevard Arago, 75014
Paris, France.\\
$^2$ School of Physics, University of Sydney, NSW 2006, Australia.\\
}
\pubyear{2003}
\begin{document}
\maketitle

\begin{abstract}
We present the first results from a pilot study to search for distant
radio galaxies in the southern hemisphere ($\delta < -32\degr$).
Within a 360\,deg$^2$ region of sky, we define a sample of 76
ultra--steep spectrum (USS) radio sources from the 843~MHz Sydney
University Molonglo Sky Survey (SUMSS) and 1.4~GHz NRAO VLA Sky Survey
(NVSS) radio surveys with $\alpha_{843}^{1400} < -1.3$ and $S_{1400} >
$15~mJy. We observed 71 sources without bright optical or
near-infrared counterparts at 1.385~GHz with the ATCA, providing
$\sim$5\arcsec\ resolution images and sub-arcsec positional accuracy.
To identify their host galaxies, we obtained near-IR $K-$band images
with IRIS2 at the AAT and SofI at the NTT.  We identify 92\% of the
USS sources down to $K\sim 20.5$.  The SUMSS--NVSS USS sources have a
surface density more than 4 times higher than USS sources selected at
lower frequencies. This is due to the higher effective selection
frequency, and the well-matched resolutions of both surveys
constructed using the same source fitting algorithm. The scattering of
$\alpha >-1.3$ sources into the USS sample due to spectral index
uncertainties can account for only 35\% of the observed USS sources.
Since our sample appears to contain a similar fraction of very distant
($z>3$) galaxies, selecting USS sources from SUMSS--NVSS should allow
us to identify large numbers of massive galaxies at high redshift.
\end{abstract}


\section{Introduction}
High redshift radio galaxies (HzRGs) provide an ideal opportunity to
study the early universe and gain insights into the formation and
evolution of massive galaxies. The well known Hubble $K-z$ relation
\citep[\eg][]{lil84,deb02} establishes that at $z\gtrsim 1$, radio
galaxies can be used to trace the most massive star-forming
populations. Thanks to the large sky coverage of present-day radio
surveys, we can find the rare sources $\gtrsim 2$ magnitudes more
luminous than similar redshift sources found in optical/near-IR
surveys.  
Together with the evolutionary path tracing HzRGs to low-redshift
($z\leq1$) elliptical galaxies \citep[\eg][]{fra98,sad2003,dun03b} and
the imperative that radio galaxies harbour central supermassive black
holes (SMBHs), HzRGs are excellent laboratories for studying the
earliest, most massive stellar systems.

HzRGs show diffuse morphologies with strong evidence supporting high
star-formation rates \citep{deb2003,wvb99b}, starkly different to
their low-redshift counterparts which are massive elliptical galaxies
with little or no star formation. However, with only 26 HzRGs known at
$z>3$, the statistics are still unreliable. Addressing this issue is a
major concern in our study.
The search for HzRGs will also constrain formation mechanisms for
SMBHs which are still very uncertain \citep[\eg][]{loeb93,dun03a}. 
However, a galaxy with a SMBH at $z>5$ already implies that if its
SMBH is not primordial, then it must form within $< 1.3~$Gyr
\citep[for $\Omega_{\rm m}=0.27, \Omega_{\Lambda}=0.73, H_0=71 $
km~s$^{-1}$~Mpc$^{-1}$;][]{spe03,ton03}.  Only one HzRG is known at
$z>5$ \citep{wvb99a}, and it is unclear whether HzRGs exist at earlier
times. Clearly, the discovery of more HzRGs will dramatically increase
our knowledge of galaxy formation and SMBH formation timescales.

Although the large optical and infrared telescopes which probe the
distant universe are increasingly concentrated in the southern
hemisphere, the search for HzRGs has until recently been limited to
the northern hemisphere because of a lack of sensitive radio imaging
surveys in the south.  This has been rectified with near completion of
the Sydney University Molonglo Sky Survey \citep[SUMSS;][]{bock99} at
843~MHz which will be used, together with its forerunner the 408~MHz
Molonglo Reference Catalogue
\citep[MRC;][]{lar81}, to search for the highest--redshift radio 
galaxies in the south.  In this paper we present the results of a
pilot study from SUMSS and the 1.4~GHz NRAO VLA Sky Survey
\citep[NVSS;][]{con98}.

Our search for HzRGs consists of four steps: (i) definition of the
sample, (ii) high-resolution radio imaging to obtain accurate positions and
morphological information, (iii) near-IR identifications of the host
galaxies, and (iv) optical and/or near-IR spectroscopy to measure
their redshifts. Here we present the results from the first three
steps of our program.

\section{Sample definition}
We used the 2001 November 23 pre-release version of the SUMSS
catalogue \citep{mau03}, and version 39 of the NVSS catalogue
\citep{con98} to construct a sample of ultra steep spectrum (USS) sources. 
Note that this is a preliminary version of the SUMSS catalogue which
has some differences (see below) from the first publicly--released
version, version 1.0 of 2003 February 25. The overlapping area with
NVSS comprises 29 SUMSS mosaics of 4\fdg3 $\times$ 4\fdg3 each.  Since
we only consider sources with $\delta <-32$\degr (see \S 3.1), the
total sky area of our USS sample is 0.11 steradians, or 360\,deg$^2$.
This area lies at $b<-31$\degr, so Galactic extinction should not be a
problem in identifying the host galaxies.

\setcounter{figure}{0}
\begin{figure}
\psfig{file=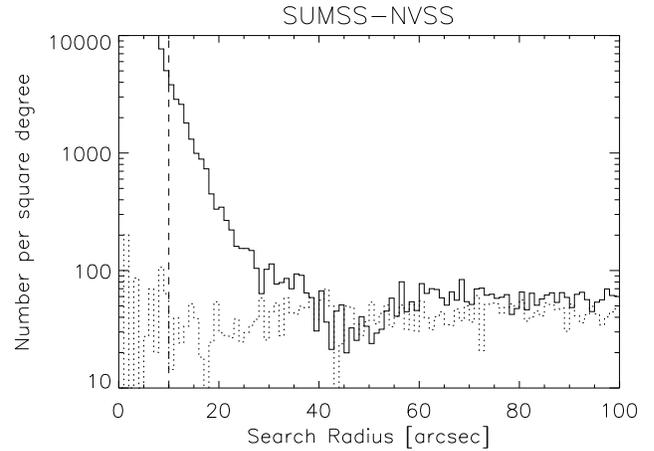,width=8.8cm}
\caption{The local density of NVSS sources around SUMSS sources as a
function of search radius. The dotted histogram represents the
distribution of random associations (see text), and flattens off near
the average source density in the NVSS survey. The vertical dashed
line is the search radius adopted for our USS sample.}
\label{searchradius}
\end{figure} 

To determine which search radius to adopt in our catalogue
correlation, we plotted the density of NVSS sources around SUMSS
sources as a function of search radius (Fig.~\ref{searchradius}). We
compared this distribution with the expected contribution from random
correlations.  To create a random position catalogue, we shifted the
NVSS positions by 1\degr\ in declination. This comparison indicates
that at search radii $>$40\arcsec, there are no more correlations than
expected from a random distribution. Because we want our USS sample to
be reliable, we used a 10\,\arcsec\ search radius. From
Fig.~\ref{searchradius}, we find that the density of NVSS sources at
offsets of 10\,\arcsec\ is $\sim$100$\times$ higher than the random
distribution, so we expect $<$1 chance coincidence between a SUMSS and
NVSS source in our sample of 76 sources. Using this cutoff, we exclude
the 2.8\% real sources with offsets $>$10\arcsec.

The difference in resolution between SUMSS (45\arcsec $\times$
45\arcsec cosec$|\delta|$) and NVSS (45\arcsec $\times$ 45\arcsec) may
cause a source to be resolved into two or more components in one
catalogue, but not in the other\footnote{This can explain the small
dip in the source density of Fig.~\ref{searchradius} between search
radii of 40\arcsec\ and 55\arcsec.}. Since this would lead to
spuriously steep spectral indices, we excluded those objects which
had another source within 100\arcsec, which corresponds to 24\% of
the matched sample.  This results in 9556 matches.  The NVSS positions
are expected to be more accurate than the SUMSS positions in this
declination zone\footnote{At dec.\ $-32$\degr, the SUMSS beam is
45\arcsec $\times$ 81\arcsec, i.e.\ significantly larger than the NVSS
beam in the N--S direction.}, so we refer to the sources in common by
their NVSS name in the standard IAU format.

We used a spectral index\footnote{Spectral index $\alpha$ is defined
by the power law $S_{\nu} \propto \nu^{\alpha}$} cutoff
$\alpha_{843}^{1400} < -1.3$, calculated using the integrated SUMSS
flux densities from the preliminary catalogue.  We also imposed a flux
limit of $S_{1400} \ge 15$~mJy to increase the accuracy of the
spectral indices. Of the 78 sources thus selected, we excluded two
(NVSS~J000920$-$351354 and NVSS~J012653$-$322807) which have bright
foreground stars in the field, as this complicates optical and near-IR
imaging.  The final sample of 76 sources is listed in
Table~2.

Note that 23 sources in our sample have spectral indices
$\alpha_{843}^{1400} > -1.3$, and nominally fall outside our
selection criterion. This is because in Table ~2, we have used the
integrated flux densities from version 1.0 of the SUMSS
catalogue. This version calculates the integrated flux densities from
the peak intensity and the widths of the gaussian fit using the same
method as described by \citet{con98}, while the preliminary version
uses the raw output from the \aips\ task {\tt VSAD} (T.\ Mauch,
private communication). We have manually checked the integrated flux
densities of all 76 sources directly from the SUMSS images, and
conclude that the values in the version 1.0 catalogue are reliable
(there is no systematic difference, and no sources deviate by more
than 25\%). The difference between the integrated flux densities
appears to be most significant for the largest radio sources (using
radio sizes determined from our ATCA imaging, see \S 3.1), confirming
that the preliminary catalogue systematically overestimated the
integrated flux density.
Such errors show up easily in USS samples, as we select the sources
with the largest differences between the SUMSS and NVSS flux
densities.  In the following, we shall therefore only use the
integrated flux densities from the version 1.0 catalogue. This implies
that our sample contains only 53 real USS sources with
$\alpha_{843}^{1400} < -1.3$. We have decided to retain the other 23
sources to provide a baseline in spectral index and radio size against
which to search for correlations with other properties (\eg\ $K-$band
magnitude, redshift) in our sample.

When using version 1.0 of the SUMSS catalogue, we find a total of 212
sources with $\alpha_{843}^{1400} < -1.3$. Of these sources, 69 are
in new fields since added to the SUMSS coverage. The remaining 143
sources are all within the same fields, and qualify as USS sources due
to the revised determination of the SUMSS flux density. Most of these
sources have just made it into the USS cutoff: the median spectral
index of the 90 'new' sources is $\bar{\alpha}_{843}^{1400} = -1.37$,
while for the 53 USS sources in this paper, $\bar{\alpha}_{843}^{1400}
= -1.56$. Because this new version of the SUMSS catalogue became
available only after our radio and near--IR observations, these 'new'
sources are not considered here.

\subsection{Literature}
We searched for obvious optical and $K-$band identifications at the
NVSS positions using the Digitized Sky Survey (DSS) and the 2 Micron
All Sky Survey \citep[2MASS;][]{skr97}. Of the five sources
detected in the DSS, three were also seen in 2MASS. We did not obtain
more accurate radio or $K-$band identifications for these sources, as
they are likely to be located at low ($z \ll 1$) redshifts. As a
result, we lack $K-$band information for two sources, and for these,
we quote the 2MASS 10$\sigma$ sensitivity limit. For the three sources
detected in 2MASS, we quote their 'total magnitudes from fit
extrapolation' and list these in column~10 of Table~2. 
Four sources have measured spectroscopic redshifts $0.15 < z < 0.26$
from the 2dF Galaxy Redshift Survey \citep[2dFGRS;][]{col01}; these 
are listed in column~13 of Table~2.

\section{Observations and data reduction}

\subsection{ATCA}
We used the Australia Telescope Compact Array (ATCA) over the period
UT 2001 December 12--14 to measure accurate radio positions and
morphologies of 71 sources in our USS sample. The 6A configuration was
used, spanning baselines from 330\,m to 6\,km.  We observed
simultaneously at 1.344~GHz and 1.432~GHz, and obtained 4--10 cuts of
3~minutes for each target, spread in hour angle.  Primary flux
calibration was based on observations of PKS~B1934$-$638 and the phase
calibrators were PKS~B1933$-$400, B2341$-$351 and B0153$-$410. To
avoid high azimuth tracking speeds, we excluded sources near the local
zenith by limiting our sample to $\delta <-32$\degr.

We used the {\tt CAONIS} on-line data-reduction tool to determine if a
source was detected with sufficient signal/noise (S/N). Once a source
had been clearly detected, and its morphology determined, we skipped
to the next source, with the aim of obtaining a set of radio maps with
a uniform S/N level and uv-coverage. Table~\ref{journal} gives the
total on-source integration times for each source.

The resulting synthesized beam widths are $\sim 9\arcsec \times
5\arcsec$, and the positional accuracy is estimated to be
$<1\arcsec$. We followed standard data reduction procedures in
{\smallaipsfont MIRIAD}, and combined the 1.344~GHz and 1.432~GHz
uv-data to increase the signal-to-noise and better sample the
uv-plane. The final images have a rms noise level of $\sim$0.4~mJy.

For approximately 30\% of our sample, positive $K-$band identifications
were confused by several sources located near the radio derived
positions. We therefore obtained follow-up radio observations to
increase the sensitivity of the previous images. We re-observed 20 USS
sources on UT 2003 July 29--30 in the 6A configuration. We
observed simultaneously at 1.384~GHz and 2.368~GHz and obtained 7--8
cuts of 4--6~minutes for each target, spread over 12 hours in hour
angle. The phase calibrators observed were PKS~B1933$-$400,
B2254$-$367, B0008$-$421 and B0153$-$410. All other parameters and
data reduction techniques are as described above.

\subsection{AAT}
We obtained $K_S$band imaging of 70 sources from our sample on the
nights of UT 2002 October 17 to 20, using the new IRIS2 instrument
\citep{gil00} at the 3.9m Anglo-Australian Telescope at Siding
Spring Observatory. Conditions were mostly photometric, but the seeing
was highly variable on some nights. This led to a loss of sensitivity
of up to 2 magnitudes when the $K_S$band seeing was 3\farcs0, as
compared to good conditions with 1\farcs0 seeing.  We used typical
integration times of 64~s, consisting of 8 co-adds of 8~s each in a
14-point random jitter pattern within a $40\arcsec \times 40\arcsec$
box. The detector is a 1024$\times$1024 HAWAII HgCdTe array, with a
pixel scale of 0\farcs446/pixel, resulting in a $\sim 8\arcmin \times
8\arcmin$ field of view.

We used the online data-reduction tool {\tt ORAC-DR} \citep{cav03} to
sky-subtract, register and sum our data. This allowed us to re-observe
several of the sources which were not detected after a first 14-point
dither pattern. This quick data reduction also allowed us to
re-observe 8 sources where we did not detect the object due to loss of
sensitivity during periods of bad seeing. In our final data reduction,
we retained only data obtained with a $K_S$band seeing $<$2.0\arcsec,
except for 6 sources with $K<18$.

\setcounter{table}{0}
\begin{table*}
\caption{Integration times for the radio and near-IR observations}
\label{journal}
\begin{center}
\begin{tabular}{lrrrlrrr}\hline
Name & ATCA & AAT & NTT & \hspace* {0.5cm} Name & ATCA & AAT & NTT \\
 & s & s & s & & s & s & s \\
\hline
NVSS~J001339$-$322445 & \nodata & \nodata & \nodata &  \hspace* {0.5cm} NVSS~J230035$-$363410 &  720    &  896    & 2400    \\ 
NVSS~J002001$-$333408 & 3600    &  896    & \nodata &  \hspace* {0.5cm} NVSS~J230123$-$364656 &  900    &  896    & 2400    \\ 
NVSS~J002112$-$321208 &  900    &  896    & \nodata &  \hspace* {0.5cm} NVSS~J230203$-$340932 & 2376    & 1792    & \nodata \\ 
NVSS~J002131$-$342225 &  900    &  896    & \nodata &  \hspace* {0.5cm} NVSS~J230404$-$372450 & 3924    &  896    & \nodata \\ 
NVSS~J002219$-$360728 & \nodata & \nodata & \nodata &  \hspace* {0.5cm} NVSS~J230527$-$360534 & 2376    & \nodata &  900    \\ 
NVSS~J002352$-$332338 & 2412    &  896    & \nodata &  \hspace* {0.5cm} NVSS~J230822$-$325027 &  900    &  896    & \nodata \\ 
NVSS~J002359$-$325756 &  900    &  896    & \nodata &  \hspace* {0.5cm} NVSS~J230846$-$334810 & 1260    &  896    & \nodata \\ 
NVSS~J002402$-$325253 &  720    &  896    & \nodata &  \hspace* {0.5cm} NVSS~J230954$-$365653 & 3924    &  896    & 1500    \\ 
NVSS~J002415$-$324102 &  900    &  896    & \nodata &  \hspace* {0.5cm} NVSS~J231016$-$363624 &  900    &  896    & \nodata \\ 
NVSS~J002427$-$325135 &  900    &  896    & \nodata &  \hspace* {0.5cm} NVSS~J231144$-$362215 & 1260    & 1792    & 2400    \\ 
NVSS~J002627$-$323653 & \nodata & \nodata & \nodata &  \hspace* {0.5cm} NVSS~J231229$-$371324 &  900    &  896    & \nodata \\
NVSS~J002738$-$323501 & 3348    &  896    & 1680    &  \hspace* {0.5cm} NVSS~J231311$-$361558 &  900    &  896    & \nodata \\ 
NVSS~J011032$-$335445 & 1080    &  896    & \nodata &  \hspace* {0.5cm} NVSS~J231317$-$352133 &  900    & 1792    & \nodata \\ 
NVSS~J011606$-$331241 & 1080    &  896    & 2400    &  \hspace* {0.5cm} NVSS~J231335$-$370609 &  900    &  896    & \nodata \\ 
NVSS~J011643$-$323415 & 3564    &  896    & \nodata &  \hspace* {0.5cm} NVSS~J231338$-$362708 & 1260    &  896    & 2400    \\ 
NVSS~J012904$-$324815 & \nodata & \nodata & \nodata &  \hspace* {0.5cm} NVSS~J231341$-$372504 & 3564    &  896    & \nodata \\ 
NVSS~J014413$-$330457 & 3240    &  896    & 2400    &  \hspace* {0.5cm} NVSS~J231357$-$372413 & 1260    &  896    & \nodata \\ 
NVSS~J014529$-$325915 &  720    & 1792    & \nodata &  \hspace* {0.5cm} NVSS~J231402$-$372925 &  900    &  896    & \nodata \\ 
NVSS~J015223$-$333833 & 3240    & 1792    & \nodata &  \hspace* {0.5cm} NVSS~J231459$-$362859 &  900    &  896    & \nodata \\ 
NVSS~J015232$-$333952 &  900    & 1792    & \nodata &  \hspace* {0.5cm} NVSS~J231519$-$342710 & 1620    &  896    & \nodata \\ 
NVSS~J015324$-$334117 & 1080    &  896    & \nodata &  \hspace* {0.5cm} NVSS~J231726$-$371443 & 1260    & 1792    & 1260    \\ 
NVSS~J015418$-$330150 & 1260    &  896    & \nodata &  \hspace* {0.5cm} NVSS~J231727$-$352606 & 2556    &  896    & 2400    \\ 
NVSS~J015436$-$333425 & 1080    &  896    & 2400    &  \hspace* {0.5cm} NVSS~J232001$-$363246 & 1800    &  896    & 1200    \\ 
NVSS~J015544$-$330633 &  900    &  896    & \nodata &  \hspace* {0.5cm} NVSS~J232014$-$375100 & 1260    &  896    & \nodata \\ 
NVSS~J021308$-$322338 & 1260    & 1792    & \nodata &  \hspace* {0.5cm} NVSS~J232058$-$365157 & 1620    & 1344    & \nodata \\ 
NVSS~J021359$-$321115 & 1260    &  896    & \nodata &  \hspace* {0.5cm} NVSS~J232100$-$360223 & 2376    &  896    & 1380    \\ 
NVSS~J021545$-$321047 & 1260    &  896    & \nodata &  \hspace* {0.5cm} NVSS~J232219$-$355816 & 2376    &  896    & 1800    \\ 
NVSS~J021716$-$325121 & 1260    &  896    & \nodata &  \hspace* {0.5cm} NVSS~J232322$-$345250 & 1440    &  896    & \nodata \\ 
NVSS~J030639$-$330432 & 1440    &  896    & \nodata &  \hspace* {0.5cm} NVSS~J232408$-$353547 & \nodata & \nodata & \nodata \\ 
NVSS~J202026$-$372823 & 1080    &  896    & \nodata &  \hspace* {0.5cm} NVSS~J232602$-$350321 & 1440    &  896    & \nodata \\ 
NVSS~J202140$-$373942 & 1908    &  896    & \nodata &  \hspace* {0.5cm} NVSS~J232651$-$370909 &  900    &  896    & \nodata \\ 
NVSS~J202518$-$355834 & 1440    & 1792    & \nodata &  \hspace* {0.5cm} NVSS~J232956$-$374534 &  720    &  896    & \nodata \\ 
NVSS~J202856$-$353709 & 1260    &  896    & \nodata &  \hspace* {0.5cm} NVSS~J233558$-$362236 & 3600    &  896    & 1980    \\ 
NVSS~J202945$-$344812 & 1080    & 1792    & 2160    &  \hspace* {0.5cm} NVSS~J233729$-$355529 &  720    &  896    & \nodata \\ 
NVSS~J204147$-$331731 &  720    &  896    & \nodata &  \hspace* {0.5cm} NVSS~J234137$-$342230 & 2160    &  896    & 1980    \\ 
NVSS~J204420$-$334948 & 1908    & 2688    & 2400    &  \hspace* {0.5cm} NVSS~J234145$-$350624 & 3000    &  896    & \nodata \\ 
NVSS~J213510$-$333703 & 2160    & 2688    & \nodata &  \hspace* {0.5cm} NVSS~J234904$-$362451 & 1260    &  896    & \nodata \\ 
NVSS~J225719$-$343954 &  900    &  576    & \nodata &  \hspace* {0.5cm} NVSS~J235137$-$362632 & 2628    &  896    & 2400    \\ 
\hline
\end{tabular}
\end{center}
\end{table*}

\subsection{NTT}
For the 20 sources not detected in our AAT/IRIS2 images, we obtained
deeper $K_S$band images on the nights of UT 2002 November 25 to 27
using the Son of Isaac (SofI) instrument \citep{moo98} at the ESO 3.5m
New Technology Telescope (NTT). Conditions were photometric with
0\farcs7 seeing.  We used typical integration times of 60~s,
consisting of 6 coadds of 10~s each in a 15 to 40-point random jitter
pattern within $40\arcsec
\times 40\arcsec$. The detector is a 1024$\times$1024 HAWAII HgCdTe
array, with a pixel scale of 0\farcs292/pixel, resulting in a $\sim
5\arcmin \times 5\arcmin$ field of view.

We reduced the data using the NOAO IRAF package. After flat-fielding,
the data were sky-subtracted, registered, and summed using the {\tt
DIMSUM} near-IR data reduction package. Because most images were
observed with good seeing, we block-replicated the pixels by a factor
of 2 before summing the individual images.

\subsection{Astrometry}
The combined images from the {\tt ORAC-DR} and {\tt DIMSUM} reductions
contain a crude astrometrical solution based on the telescope
pointings. We used the SkycatGAIA tool to identify the non-saturated
($12<R<18$) stars from USNO-A2.0 catalogue \citep{mon98}, and
fine-tuned our astrometry accordingly. Because of the large field of
view, we could identify on average $\sim$90 stars in each IRIS2 image,
and $\sim$40 in the SofI images, allowing an accurate astrometrical
solution including the rotation and stretch terms. The main
uncertainty in the {\it relative} near-IR to radio astrometry stems
from the uncertainty in the USNO-A2.0 catalogue, which is $1\sigma
\approx 0\farcs25$ \citep{deu99}. This is more accurate than our ATCA
astrometry, and should therefore be sufficient to identify the
$K_S$band counterparts of our USS sources.

\subsection{Photometry}
We calibrated the photometry using short observations of standard
stars from the UKIRT faint standard list \citep{haw01}. This procedure
yielded typical zeropoints of ${K_S}_0=22.20 \pm 0.10$ for IRIS2, and
${K_S}_0=22.11 \pm 0.02$ for SofI (for 1 count/second, integrated over
the source). Because our resulting images have a large field of view,
they generally contain several stars which are also detected in the
DEep Near Infrared Survey of the Southern Sky
\citep[DENIS][]{epc97}. We extracted all the DENIS stars in our images
with reliable $K-$band photometry (confidence coefficient $>$75) from
the catalogue constructed at the Paris Data Analysis Center (Guy
Simon, private communication). A comparison with our photometry yields
a small systematic offset of $-0.06$~mag in both the IRIS2 and SofI
photometric zeropoints. We applied this correction to our
photometry. We estimate the uncertainty in the zeropoints by comparing
field stars observed by both telescopes, and find $\Delta({K_S}_0)
\simeq 0.05$.

We did not correct for airmass variations, because most of our objects
were observed with airmasses $<$1.7, and the airmass dependence in
$K_S$band is small compared to the fitting errors described
below. Similarly, the effect of Galactic extinction is expected to be
negligible in our fields (located at $b<-31$\degr), and no correction
has been applied. Note that all our near--IR observations used the
$K_S$-band filter and not the standard $K-$band filter. We did not
correct our magnitudes to $K-$band because we do not have any colour
information, and the correction is expected to be much smaller than
the photometric uncertainties. In the following, we shall for refer to
the magnitudes as $K-$band to facilitate comparison with the literature.

We used the IRAF task {\tt phot} to measure the magnitudes of the
$K-$band identifications. To facilitate comparison with the
literature, we use three different apertures, with diameters of
2\farcs0, 4\farcs0, and 8\farcs0. The uncertainties quoted are our
best estimates, which include both the zero-point uncertainty and the
fitting uncertainty given by the {\tt phot} routine.

\section{Results}
Table~2 lists our sample with the results from our radio and near-IR
imaging. The columns are:

\begin{description}
\item[(1)]{Name of the source in IAU J2000 format.}
\item[(2)]{The integrated 843~MHz flux density from the SUMSS
catalogue.}
\item[(3)]{The integrated 1.4~GHz flux density from the NVSS catalogue.}
\item[(4)]{The two-point spectral index between SUMSS and NVSS.}
\item[(5)]{Largest angular size, as determined from the ATCA maps. 
For single component sources, this is the de-convolved major axis of
the elliptical Gaussian, or, for unresolved sources (preceded by $<$),
an upper limit is given by the resolution. For multiple component
sources, this is the largest possible separation between their
components.}
\item[(6)]{De-convolved position angle of the radio structure, as determined 
from the ATCA maps, measured North through East. For multiple
component sources, this is the orientation of the most widely
separated components used to calculate the LAS (col. 5).}
\item[(7)]{The telescope used to obtain the $K-$band imaging.}
\item[(8)--(10)]{$K-$band magnitude determined in apertures with 
diameters of 2\arcsec, 4\arcsec, and 8\arcsec. For the 2MASS
photometry, these are the total magnitudes from fit extrapolation.}
\item[(11)--(12)]{J2000 coordinates of the radio source, either
measured from the ATCA images or from the NVSS catalogue, as indicated
in column 13. The positions in the ATCA maps have been fitted with a
single two-dimensional elliptical Gaussian. For multiple component
sources, the geometric midpoint is given, unless mentioned otherwise
in the notes on individual sources.}
\item[(13)]{Origin of the radio position given in columns 11 and 12:
A=ATCA image, N=NVSS catalogue.}
\item[(14)--(15)]{J2000 coordinates of the $K-$band identification.}
\item[(16)]{The spectroscopic redshift from the 2dFGRS.}
\end{description}

Notes on individual sources are given in the Appendix.
Figure~\ref{overlays} shows overlays of the ATCA 1.4\,GHz or 2.4~GHz
radio contours onto $K-$band images.

\section{Discussion}

\setcounter{figure}{2}
\begin{figure}
\psfig{file=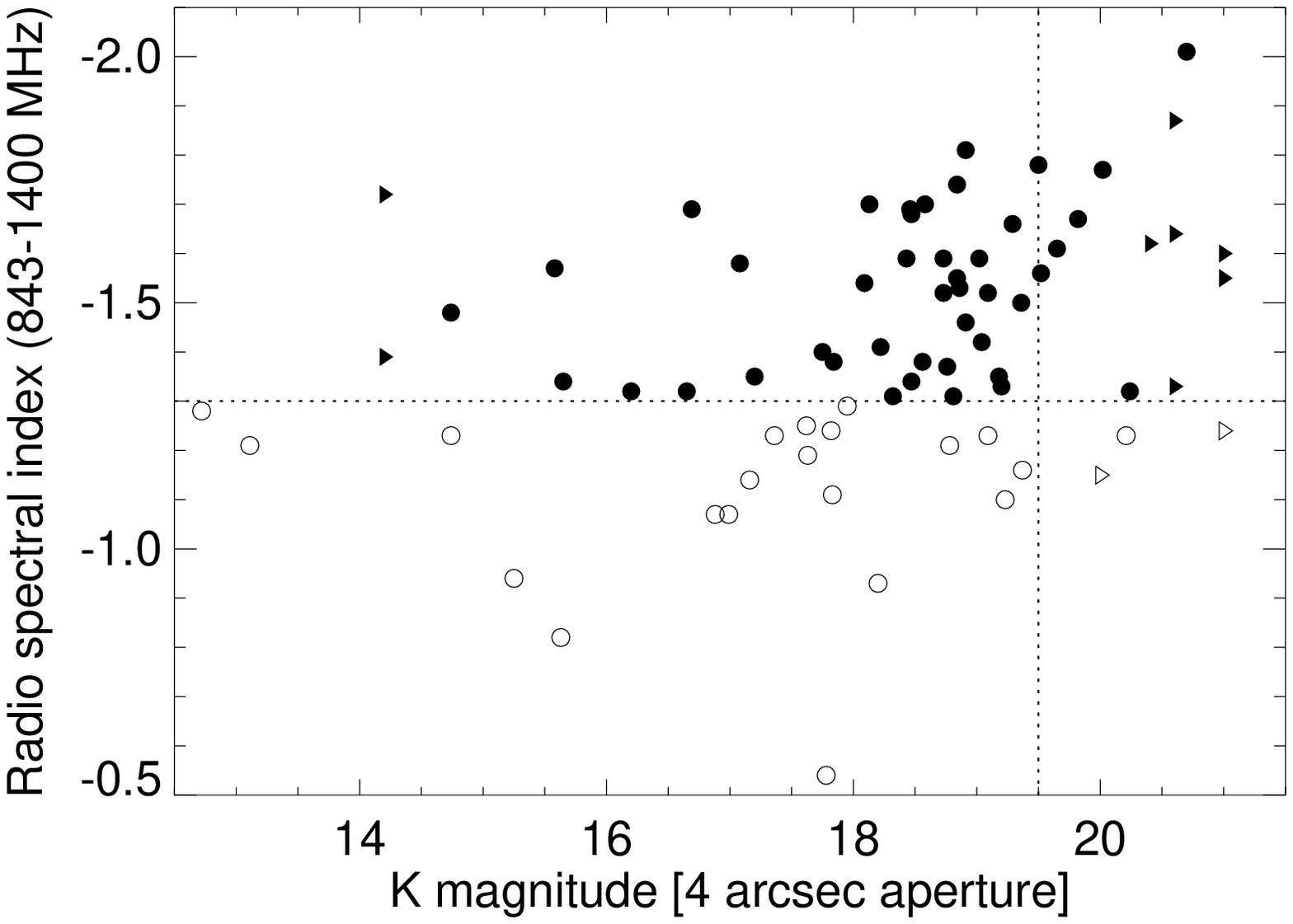,width=8.8cm}

\caption{Relation between K magnitude (in a 4\,arcsec aperture) and
the radio spectral index between 843\,MHz and 1.4\,GHz. Triangles show
upper limits. Filled symbols represent sources with
$\alpha_{843}^{1400} < -1.3$, and open symbols those with
$\alpha_{843}^{1400}>-1.3$. The vertical dotted line indicates the
expected $K-$band magnitude of a $z\sim$3 source, and the horizontal
dotted line our adopted spectral index cutoff.}
\label{Kalpha}
\end{figure}

\setcounter{figure}{3}
\begin{figure}
\psfig{file=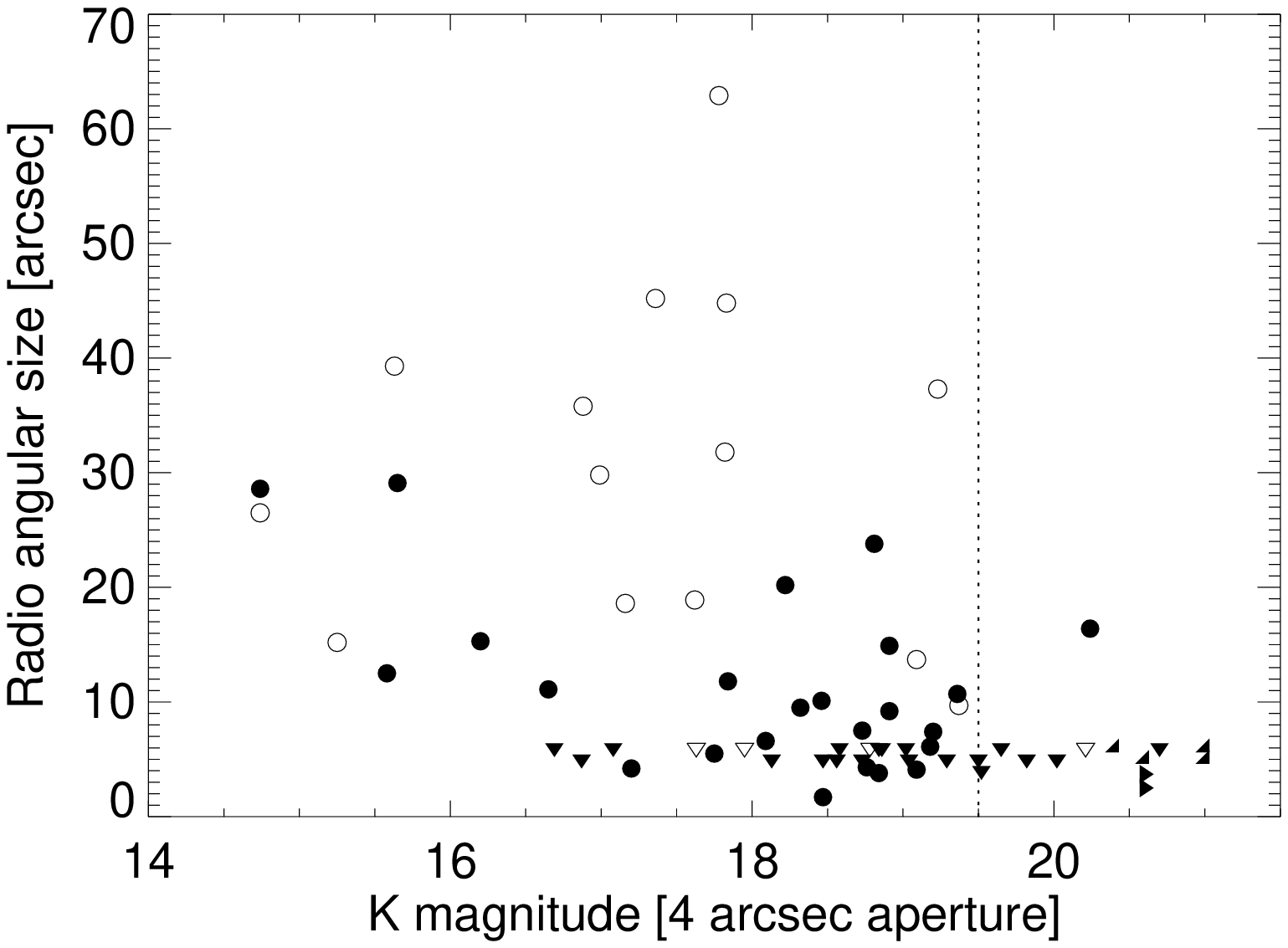,width=8.8cm}

\caption{Relation between K magnitude (in a 4\,arcsec aperture) of a
radio galaxy and the angular size of the radio source. Symbols are as
in Fig.~\ref{Kalpha}. The vertical dotted line indicates the expected
$K-$band magnitude of a $z\sim$3 source. Note the paucity of large
radio sources with $K>19.5$.}
\label{Klas}
\end{figure} 

\subsection{Correlations}
We now examine the correlations between the different source
parameters in our sample. Because 30\% of our sample contains sources
with $\alpha_{843}^{1400} > -1.3$ and large radio sizes, we can study
the dependence of $K-$band magnitude on these
parameters. Fig.~\ref{Kalpha} shows that the $K-$band magnitude and
radio spectral index are correlated. The generalized Spearman rank and
generalized Kendall's tau coefficients \citep[\eg][]{iso86} suggest
this relation is significant at the 99.95\% level. This correlation
arises because both radio spectral index and $K-$band magnitude are
correlated with redshift, reinforcing our approach of using $K-$band
magnitude as an additional criterion for selecting the highest
redshift candidates.

Several samples of USS sources \citep{blu98,ste99} have applied a
radio angular-size cutoff to exclude large radio sources, which are
thought to be at relatively low redshift. We can check the efficiency
of this approach with our sample, as we have radio size information
for 95\% of our sources, and $K-$band information to estimate their
redshifts (see \S 5.3). Note that our sample also has an implicit
angular size cutoff due to the exclusion of SUMSS sources with more
than one NVSS source within 100\arcsec (see \S 2).  Considering only
the 53 sources with $\alpha_{843}^{1400} < -1.3$ and radio size
information, there indeed appears to be a correlation between radio
size and $K-$band magnitude (see Fig.~\ref{Klas}). The generalized
Spearman rank and Kendall's tau coefficients suggest that this
correlation is significant at the 99.7\% level. Because our sample
contains only 29 radio sources larger than 10\arcsec\ (only 12 of
which have $\alpha_{843}^{1400} < -1.3$), we consider this suggestive,
but not conclusive evidence that the most distant radio sources in our
radio sample have smaller radio sizes. Nevertheless, it is remarkable
that only two sources with $K(4\arcsec)>19.5$ have a radio size larger
than 6\arcsec. This suggests that a significant fraction of the
compact steep spectrum sources in our sample are either at high
redshift or are heavily obscured by dust \citep{reu03}.

\subsection{Surface density of USS sources}

In this paper, we select USS sources on the basis of the radio
spectral index between 843 and 1400\,MHz.  We now evaluate how well
this works in comparison with the techniques used in earlier USS
samples.  There are two considerations here: the {\it number}\ of
sources which this technique yields, and the {\it fraction} of USS
sources which lie at high redshift.  

Table~\ref{surfacedensity} compares the results from our SUMSS--NVSS
pilot study with earlier USS samples observed by \citet{deb02} and
\citet{jar01}. 
The SUMSS--NVSS USS selection yields a USS surface density more than
four times higher than the WENSS--NVSS sample of De Breuck et al.\
(2002) at the same 1.4\,GHz flux level. Note that due to the
incompleteness of our USS sample (see \S 2), these numbers are
strictly lower limits. There are several explanations for this higher
density:

First, the higher {\it observed} selection frequency of 843~MHz
compared to 325~MHz means that we are also sampling a higher {\it
rest-frame} spectral index of the radio sources. Because most radio
spectra of powerful radio galaxies and compact steep spectrum sources
have a tendency to steepen towards higher frequencies
\citep[\eg][]{gop88,man95,ath98,mur99,blu99,deb00,sohn03}, we are effectively
probing an intrinsically steeper part of the radio
spectrum. Therefore, on average, $\alpha_{843}^{1400}=-1.3$
corresponds to $\alpha_{325}^{1400}>-1.3$. This slightly more relaxed
cutoff leads to a fast increase in the number of USS selected sources
(due to the steep tail of the spectral index distribution), but due to
the other two effects described here, we cannot use this to estimate
the $\alpha_{843}^{1400}$ cutoff equivalent to
$\alpha_{325}^{1400}=-1.3$.

Second, the SUMSS and NVSS catalogues have both been constructed with
the \aips\ task {\tt VSAD} \citep{con98,mau03}, while the WENSS survey
uses an IDL-based fitting routine \citep{ren97}. Because both USS
samples consider only isolated sources in both catalogues, the use of
the same fitting routine in both surveys is more appropriate. For
example, \citet{deb00} considered only `single component' WENSS
sources, which may well exclude a substantial fraction of real USS
sources. Furthermore, the spatial resolution of SUMSS (45\arcsec
$\times$ 45\arcsec cosec$|\delta|$) is better matched to NVSS
(45\arcsec $\times$ 45\arcsec) than that of WENSS (54\arcsec $\times$
54\arcsec cosec$|\delta|$), so fewer sources will be removed by our
selection criterion to exclude sources resolved in only one of the two
catalogues (see \S 2). The WENSS--NVSS USS sample is therefore less
complete than the SUMSS--NVSS USS sample.

Third, the shorter frequency baseline between SUMSS and NVSS leads to
a median uncertainty in the derived USS spectral indices
$\Delta\bar{\alpha}_{843}^{1400} = 0.12$, while for WENSS-NVSS, this
is $\Delta\bar{\alpha}_{325}^{1400} = 0.04$ \citep{deb00}. Because the
$\alpha < -1.3$ cutoff is on the very steep tail of the spectral
index distribution, we expect more sources with $\alpha>-1.3$ to
scatter into into the USS sample than there will be $\alpha<-1.3$
scattering out. To estimate the magnitude of this effect, we have
generated a random sample of sources drawn from the SUMSS--NVSS
spectral index distribution, and added random spectral index
uncertainties to this \citep{ren99}. Here, we use the publicly
released version of the SUMSS catalogue, containing 212 USS sources
with $\alpha<-1.3$ to get reliable estimated of this effect. Adopting
a mean spectral index $\bar{\alpha}=-0.82$ and a standard deviation of
$\sigma_{\alpha}=0.25$, we expect that due to uncertainties in the
spectral indices in our sample, $\sim$40 real $\alpha<-1.3$ have
observed $\alpha>-1.3$, while $\sim$115 real $\alpha>-1.3$ have
observed $\alpha<-1.3$. Hence, this would lead to a surplus of
$\sim$75 sources (35\%) in a USS sample \citep[compared to $\sim$3\%
in the WENSS--NVSS sample; ][]{deb00}. However, we do not expect that
this `contamination' of $\alpha > -1.3$ sources will decrease the
fraction of $z>3$ radio galaxies in our SUMSS--NVSS USS sample because
the scatter in the $\alpha - z$ relation is quite large.

Table~\ref{surfacedensity} also shows that the fraction of USS sources
which have $K-$band IDs fainter than $K_{\rm 64~kpc}>$19.5\,mag,
making them candidates for very distant ($z>3$) radio galaxies,
appears similar in the SUMSS--NVSS and WENSS--NVSS USS sample with a
$S_{1400}>$15~mJy cutoff.  However, the $K-$band photometry of the
WENSS--NVSS USS sample is not complete as most of the WENSS--NVSS
sources observed in $K-$band were pre-selected to be undetected in
optical imaging ($R \simgt 24$). This preselection will clearly have
removed a substantial number of `intermediate redshift' sources with
$K<19.5$. If the 20 $R-$band detected sources not observed at $K-$band
are all assumed to have $K<19.5$, then the fraction decreases to
19.6~\%, slightly lower than the value found for the SUMSS-NVSS sample.

Thus it appears that spectral index selection between 843 and
1400\,MHz is an efficient way of finding distant galaxies, and the
full SUMSS--NVSS USS sample should be capable of finding large numbers
of massive galaxies at $z>3$.  Spectroscopy is clearly needed to
confirm that the $K-z$ relation holds for these galaxies.

\setcounter{table}{2}
\begin{table*}
\begin{minipage}{170mm} 
\caption{Surface density of USS sources for different selection methods. Note that these are strictly lower limits due to the incompletenesses in the various USS samples (see text).\label{surfacedensity}}
\begin{tabular}{@{}lccccccll}\hline
Sample  & Flux limit & Spectral index & Area  & Sources & USS density & Fraction of USS & Density  (sr$^{-1}$) & Ref. \\
        & (mJy)          & limit & sr    &     & (sr$^{-1}$) & with K$>$19.5 mag &  
USS$+$K$>$19.5 & \\
\hline
WENSS--NVSS & $S_{1400}>$ 10 & $\alpha_{325}^{1400}<-1.3$  & 2.27  & 343  &  151 &  12/44 (27\%) &  41$\pm12$ & 1 \\
TEXAS--NVSS & $S_{1400}>$ 10 & $\alpha_{365}^{1400}<-1.3$  & 5.58  & 268  &   48 &   8/24 (33\%) &  16$\pm6$  & 1 \\
MRC--PMN    & $S_{408}>$ 700 & $\alpha_{408}^{4800}<-1.3$  & 2.23  &  58  &   26 &   0/19        &  $<$1.4    & 1 \\
6C$^*$      & $S_{151}>$ 960 & $\alpha_{151}^{4850}<-0.981$ & 0.133 &  24  &  180 &   2/24 ( 8\%) &  15$\pm11$ & 2 \\
&&&&&&& \\		       
SUMSS--NVSS & $S_{1400}>$ 15 & $\alpha_{843}^{1400}<-1.3$  & 0.11  &  53  &  482 &  13/53 (25\%) & 118$\pm33$ & 3 \\
WENSS--NVSS & $S_{1400}>$ 15 & $\alpha_{325}^{1400}<-1.3$  & 2.27  & 233  &  103 &  11/36 (31\%) &  32$\pm10$ & 1 \\
\hline
\end{tabular}

REFERENCES: (1) \cite{deb00,deb02}; (2) \cite{blu98}, \citet{jar01}; (3) this paper.
\end{minipage}
\end{table*}

\subsection{Expected redshift distribution}
We have spectroscopic redshifts of only four sources from the
2dFGRS. To estimate the redshift distribution of the other 71 sources,
we use the Hubble $K-z$ diagram. We have fitted a linear relation to the
64~kpc radio-galaxy magnitudes in Fig.~7 of \citet{deb02}, yielding
$K=4.633\log_{10}(z)+17.266$. To calculate the 64~kpc metric
apertures, we used the average correction for $z>1$, viz. $K_{\rm 64
kpc}=K(8\arcsec)+0.2$. Figure~\ref{zexpected} shows the predicted
redshift distribution of our sample. The median predicted redshift is
1.75, which is slightly lower than for the WENSS--NVSS and Texas--NVSS
USS samples \citep{deb02}. However, when the optical preselection in
the former (see previous section) is taken into account, the
SUMSS--NVSS sample has at least as high a median expected redshift.
\setcounter{figure}{4}
\begin{figure}
\vspace*{8cm}
\includegraphics{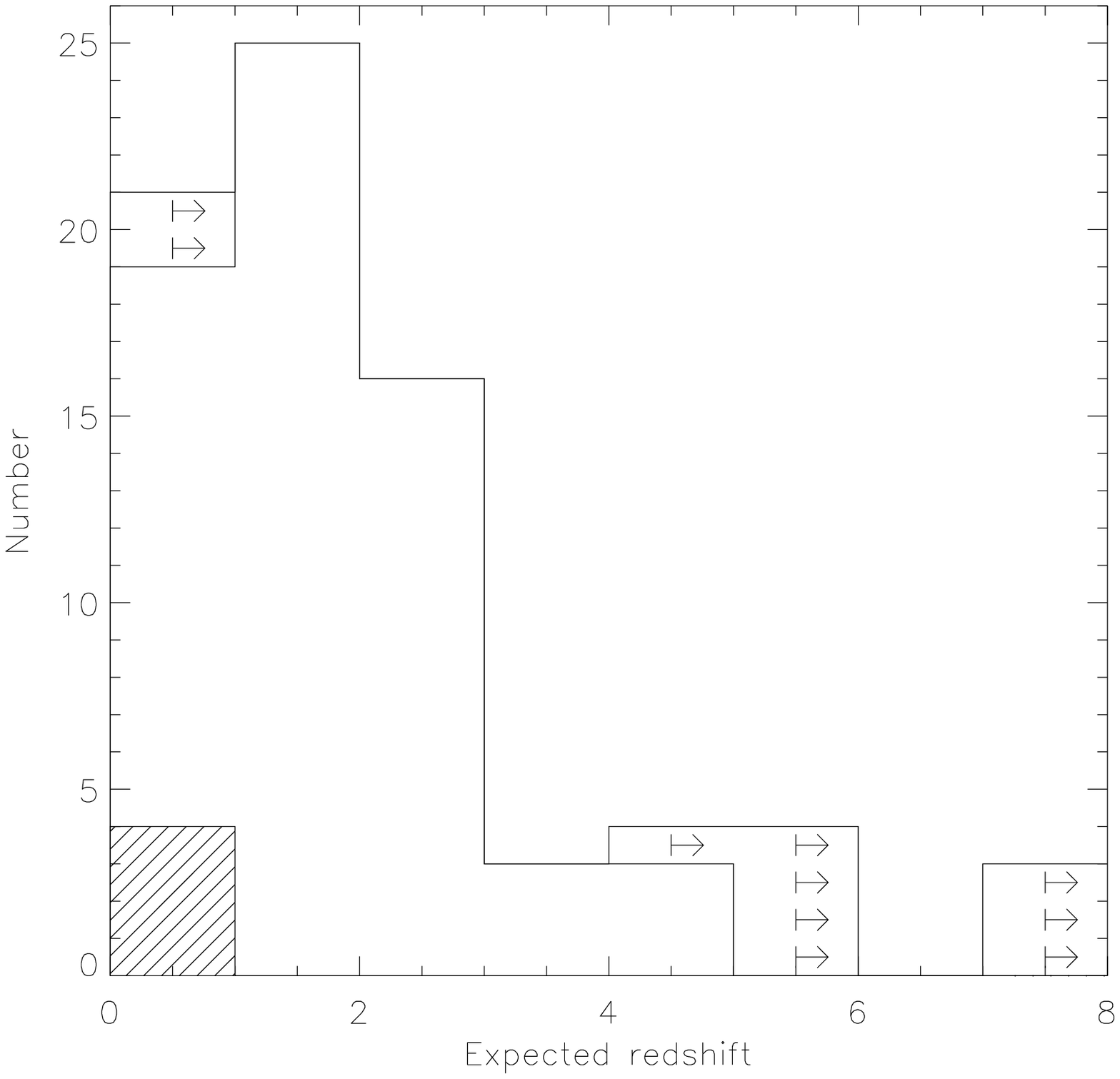}

\caption{Predicted redshift distribution of the USS sources, estimated
from the Hubble $K-z$ diagram. The shaded histogram shows the four
sources with spectroscopic redshifts from the 2dFGRS}
\label{zexpected}
\end{figure} 

\section{Conclusions}
We have constructed a sample of 76 southern USS sources from the SUMSS
and NVSS, including 53 sources with $\alpha_{843}^{1400} < -1.3$. Our
$\sim$5\arcsec\ resolution radio maps allow the identification of 92\%
of these sources down to $K\sim 20.5$. The surface density of this
SUMSS--NVSS USS sample is more than four times higher than earlier USS
samples, which can be explained by a spectral index cutoff which is
effectively more relaxed due to the higher selection frequency, and
by the use of the same source finding algorithm in both catalogues. The
higher uncertainties in the spectral indices due to the shorter
frequency baseline account for a further 35\% of $\alpha > -1.3$
source which get scattered into the USS sample.

The distribution of the $K-$band magnitudes suggests that our sample
will be at least as efficient in finding $z>3$ radio galaxies as
earlier USS samples. We intend to obtain full spectroscopic redshift
information for this sample to increase the number of potential HzRG
targets for follow-up studies with 8m-class telescopes in the southern
hemisphere, and to study the radio power dependence in the Hubble
$K-z$ diagram out to $z>3$.

By the end of 2003, SUMSS is expected to cover most of the $\delta <
-30$\degr\ region. When combined with the re-analysis of the 408~MHz
MRC (Crawford, in preparation), this will allow the construction of a
sensitive USS sample at $\delta < -40$\degr, a region which is
virtually unexplored in extragalactic radio astronomy.

\section{Acknowledgements}
We thank the referee Philip Best for his valuable comments, Tom Mauch
for useful discussions and Guy Simon for providing the $K-$band data
from the DENIS survey.  This publication makes use of data products
from the Two Micron All Sky Survey, which is a joint project of the
University of Massachusetts and the Infrared Processing and Analysis
Center/California Institute of Technology, funded by the National
Aeronautics and Space Administration and the National Science
Foundation.  This work was supported by PICS/CNRS (France) and
IREX/ARC (Australia), and by a Marie Curie Fellowship of the European
Community programme `Improving Human Research Potential and the
Socio-Economic Knowledge Base' under contract number
HPMF-CT-2000-00721.

{}


\begin{figure*}
\vspace*{-1cm}
\psfig{file=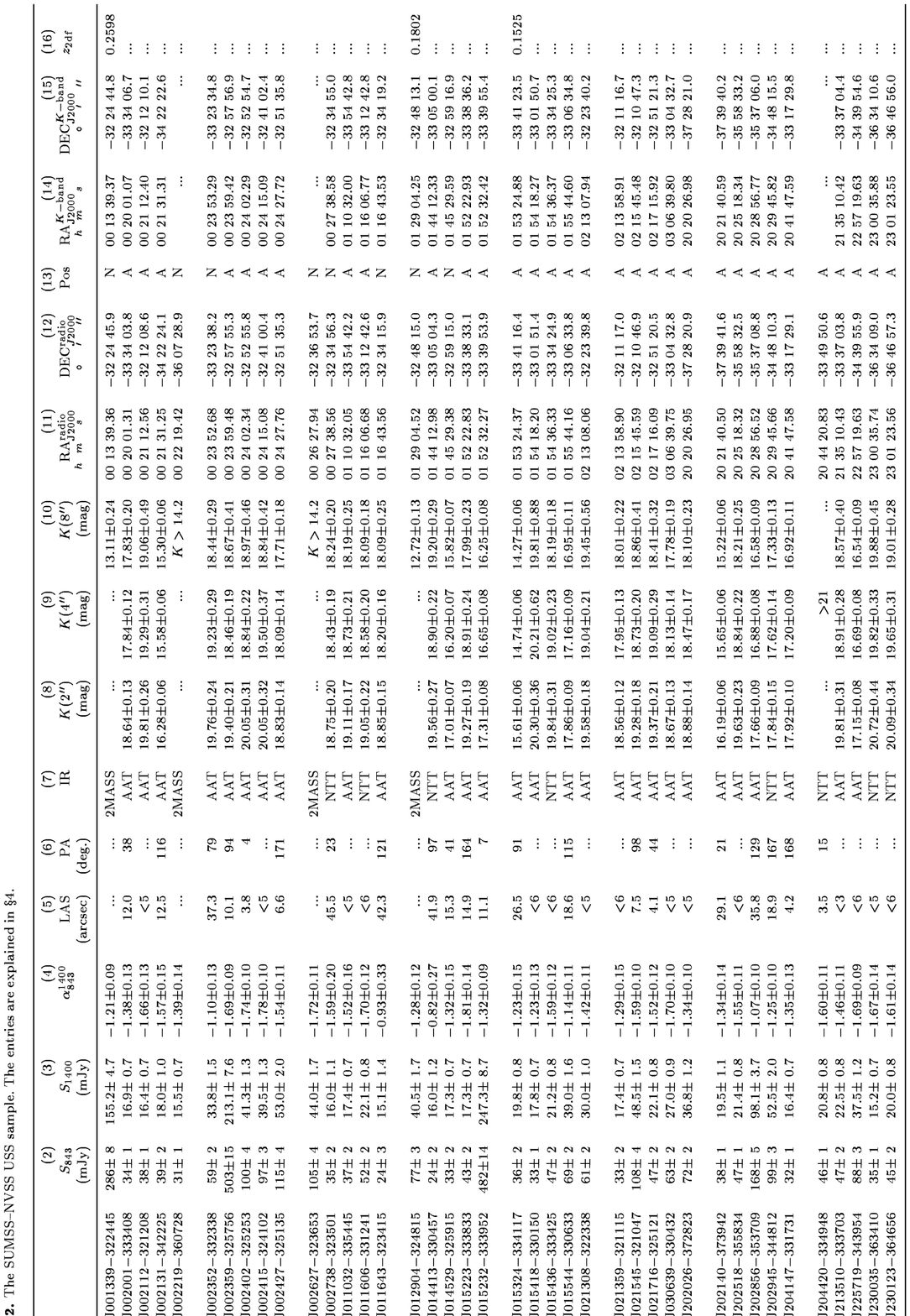,height=26cm,angle=180}
\label{datatable}
\end{figure*}
\vspace*{-1cm}
\begin{figure*}
\psfig{file=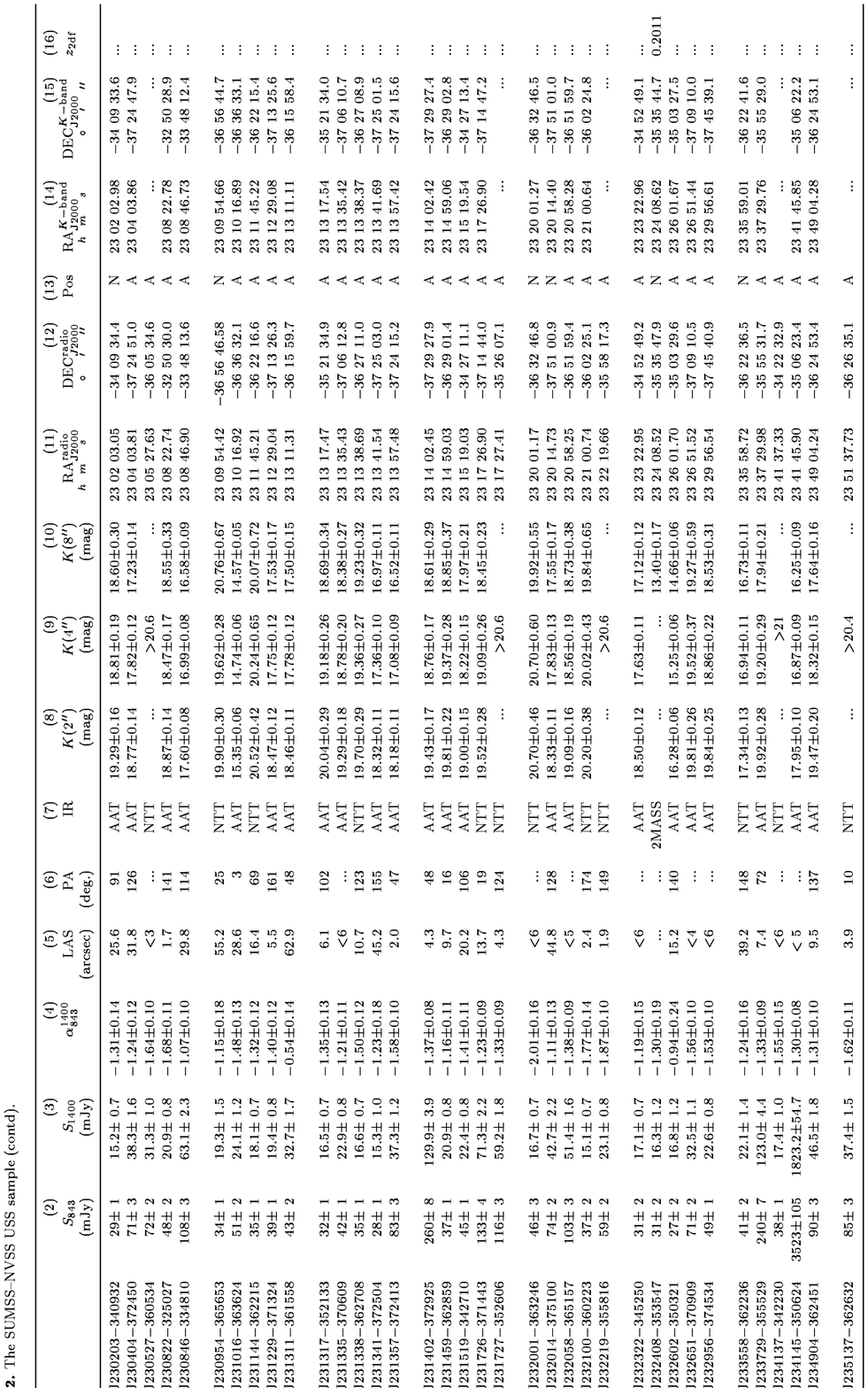,height=26cm,angle=180}
\end{figure*}

\setcounter{figure}{1}
\begin{figure*}
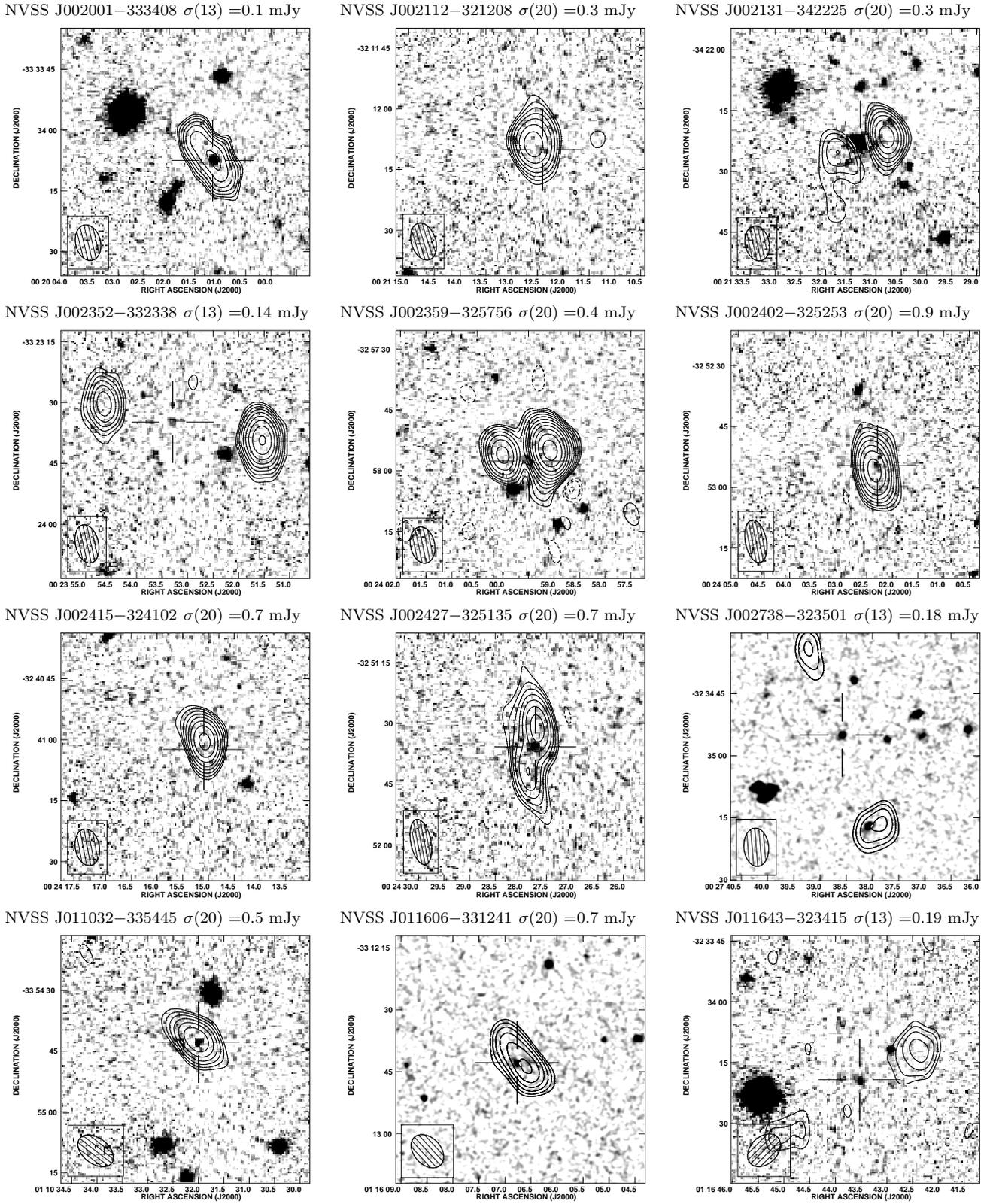

\begin{tabular}{lll}
NVSS~J002001$-$333408 $\sigma(13)=$0.1~mJy & NVSS~J002112$-$321208 $\sigma(20)=$0.3~mJy & NVSS~J002131$-$342225 $\sigma(20)=$0.3~mJy \\
\psfig{file=overlays/NVSSJ002001-333408.PS,width=5.5cm,angle=-90}&
\psfig{file=overlays/NVSSJ002112-321208.PS,width=5.5cm,angle=-90}&
\psfig{file=overlays/NVSSJ002131-342225.PS,width=5.5cm,angle=-90}\\
NVSS~J002352$-$332338 $\sigma(13)=$0.14~mJy & NVSS~J002359$-$325756 $\sigma(20)=$0.4~mJy & NVSS~J002402$-$325253 $\sigma(20)=$0.9~mJy \\
\psfig{file=overlays/NVSSJ002352-332338.PS,width=5.5cm,angle=-90}&
\psfig{file=overlays/NVSSJ002359-325756.PS,width=5.5cm,angle=-90}&
\psfig{file=overlays/NVSSJ002402-325253.PS,width=5.5cm,angle=-90}\\
NVSS~J002415$-$324102 $\sigma(20)=$0.7~mJy & NVSS~J002427$-$325135 $\sigma(20)=$0.7~mJy & NVSS~J002738$-$323501 $\sigma(13)=$0.18~mJy \\
\psfig{file=overlays/NVSSJ002415-324102.PS,width=5.5cm,angle=-90}&
\psfig{file=overlays/NVSSJ002427-325135.PS,width=5.5cm,angle=-90}&
\psfig{file=overlays/NVSSJ002738-323501.PS,width=5.5cm,angle=-90}\\
NVSS~J011032$-$335445 $\sigma(20)=$0.5~mJy & NVSS~J011606$-$331241 $\sigma(20)=$0.7~mJy & NVSS~J011643$-$323415 $\sigma(13)=$0.19~mJy \\
\psfig{file=overlays/NVSSJ011032-335445.PS,width=5.5cm,angle=-90}&
\psfig{file=overlays/NVSSJ011606-331241.PS,width=5.5cm,angle=-90}&
\psfig{file=overlays/NVSSJ011643-323415.PS,width=5.5cm,angle=-90}\\
\end{tabular}

\caption{Part of the overlays of ATCA 13cm or 20cm maps on AAT/IRIS2 or 
NTT/SofI $K-$band images. We have smoothed the SofI images using a
circular 0\farcs7 Gaussian, corresponding to the average seeing. The
contour scheme is a geometric progression in $\sqrt 2$, which implies
a factor two change in surface brightness every 2 contours. The first
contour level, indicated above each plot, is at $3\sigma_{\rm rms}$,
where $\sigma_{\rm rms}$ is the rms noise measured around the
sources. The wavelength of the radio map is given in brackets. The
restoring beams are indicated in the lower left corner of the
plots. The open cross indicates the $K-$band identification, as listed
in Table~2.  See http://www.eso.org/$\sim$cbreuck/papers.html for a version
with all 71 overlays.}
\label{overlays}
\end{figure*}

\bigskip
\appendix 
\section{Notes on individual sources}

{\bf NVSS~J001339$-$322445}: This is the northern member of a pair of
radio sources separated by 2.2\,arcmin on the sky and previously
catalogued as PKS\,0011-327. While this could be a wide double, we
consider it more likely that the two sources are unrelated as there is
no plausible optical or near-IR counterpart located between the two
sources.  The northern (steep--spectrum) source NVSS~J001339$-$322445
is associated with a bright galaxy seen on both the DSS and 2MASS
images, and so was not observed with the ATCA or AAT.  The southern
source, NVSS~J001338$-$322657, has no obvious optical counterpart.
Its flux density is 106\,mJy in NVSS and 141\,mJy in SUMSS, giving a
spectral index of $-0.56$.

\noindent
{\bf NVSS~J002001$-$333408}: The $K-$band identification is located
at the centre of the extended radio source.

\noindent
{\bf NVSS~J002112$-$321208}: The diffuse $K-$band identification is
offset by ($\Delta \alpha, \Delta \delta)$=(2\farcs0W, 1\farcs5S) from
the ATCA position.

\noindent
{\bf NVSS~J002131$-$342225}: The source is identified with the
(uncatalogued) bright galaxy located between the radio lobes.

\noindent
{\bf NVSS~J002219$-$360728}: An optical galaxy is clearly visible on
the digitized sky survey images at the NVSS position, so this source
was not observed with the ATCA or AAT.

\noindent
{\bf NVSS~J002352$-$332338}: We identify the host galaxy with the
faint $K-$band object located along the radio axis, offset by
($\Delta \alpha, \Delta \delta)$=(7\farcs6E, 3\farcs4N) from the NVSS
position.

\noindent
{\bf NVSS~J002359$-$325756}: The $K-$band identification is located
in between the radio lobes. This source is also known as
PMN~J0024$-$3258.

\noindent
{\bf NVSS~J002402$-$325253}: The diffuse $K-$band identification is
offset by ($\Delta \alpha, \Delta \delta)$=(0\farcs6W, 1\farcs1N) from
the ATCA position.

\noindent
{\bf NVSS~J002415$-$324102}: The diffuse $K-$band identification is
located at the ATCA position.

\noindent
{\bf NVSS~J002427$-$325135}: The $K-$band identification is located
in between the radio lobes.

\noindent
{\bf NVSS~J002627$-$323653}: An optical galaxy is clearly visible on
the digitized sky survey images at the NVSS position, so this source
was not observed with the ATCA or AAT.

\noindent
{\bf NVSS~J002738$-$323501}: The diffuse $K-$band identification is
located at the midpoint of the radio lobes.

\noindent
{\bf NVSS~J011032$-$335445}: The $K-$band identification is located
at the ATCA position.

\noindent
{\bf NVSS~J011606$-$331241}: The $K-$band identification is located
at the ATCA position.

\noindent
{\bf NVSS~J011643$-$323415}: We identify the compact $K-$band source
offset by ($\Delta \alpha, \Delta\delta)$=(0\farcs5W, 3\farcs6S) from
the NVSS position.

\noindent
{\bf NVSS~J012904$-$324815}: This bright source is clearly detected in
2MASS using the NVSS astrometry. We therefore did not observe it with
the ATCA or AAT.

\noindent
{\bf NVSS~J014413$-$330457}: This radio source has a complex
morphology. We identify a diffuse $K-$band source at the position of
the brightest radio component, which is most likely the core. In
table~2, we list the position of this radio core.

\noindent
{\bf NVSS~J014529$-$325915}: The diffuse radio source appears to lie
within a small a cluster of galaxies. We identify the bright
$K-$band source offset by ($\Delta \alpha, \Delta
\delta)$=(2\farcs6E, 2\farcs0S) from the ATCA position.

\noindent
{\bf NVSS~J015223$-$333833}: The $K-$band identification is located
along the radio axis, offset by ($\Delta \alpha, \Delta
\delta)$=(1\farcs6E, 2\farcs7S) from the midpoint between the radio
lobes in the ATCA image.

\noindent
{\bf NVSS~J015232$-$333952}: The bright $K-$band identification is
located between the main two radio lobes. The radio source, also known
as PMN~J0152$-$3340, has a complex morphology with a fainter lobe to
the east of the main two radio lobes.

\noindent
{\bf NVSS~J015324$-$334117}: The diffuse radio source appears to lie
within a small a cluster of galaxies. We identify the bright
$K-$band source offset by ($\Delta \alpha, \Delta
\delta)$=(6\farcs4E, 6\farcs7S) from the ATCA position.

\noindent
{\bf NVSS~J015418$-$330150}: The faint $K-$band identification is
located at the ATCA position.

\noindent
{\bf NVSS~J015436$-$333425}: The diffuse $K-$band identification is
located at the ATCA position.

\noindent
{\bf NVSS~J015544$-$330633}: The bright $K-$band identification is
located between the radio lobes.

\noindent
{\bf NVSS~J021308$-$322338}: The $K-$band identification is located
at the ATCA position.

\noindent
{\bf NVSS~J021359$-$321115}: The $K-$band identification is located
at the ATCA position.

\noindent
{\bf NVSS~J021545$-$321047}: The $K-$band identification is located
along the radio axis, offset by ($\Delta \alpha, \Delta
\delta)$=(1\farcs4W, 0\farcs2S) from the midpoint between the radio
lobes in the ATCA image.

\noindent
{\bf NVSS~J021716$-$325121}: The $K-$band identification is located
between the radio lobes.

\noindent
{\bf NVSS~J030639$-$330432}: The $K-$band identification is located
at the ATCA position. This source is also known as TXS~0304$-$332.

\noindent
{\bf NVSS~J202026$-$372823}: The $K-$band identification is located
at the ATCA position.

\noindent
{\bf NVSS~J202140$-$373942}: The $K-$band identification is located
at the central of the 3 radio components.

\noindent
{\bf NVSS~J202518$-$355834}: The $K-$band identification is located
at the ATCA position.

\noindent
{\bf NVSS~J202856$-$353709}: The bright $K-$band identification is
located between the radio lobes.

\noindent
{\bf NVSS~J202945$-$344812}: The $K-$band identification is located
along the radio axis, offset by ($\Delta \alpha, \Delta
\delta)$=(1\farcs7E, 5\farcs0S) from the midpoint between the radio
lobes in the ATCA image.

\noindent
{\bf NVSS~J204147$-$331731}: The bright $K-$band identification is
located between the radio lobes.

\noindent
{\bf NVSS~J204420$-$334948}: No $K-$band source is seen near the
radio position in our medium deep SofI image.

\noindent
{\bf NVSS~J213510$-$333703}: The faint $K-$band identification is
located at the ATCA position.

\noindent
{\bf NVSS~J225719$-$343954}: The bright $K-$band identification is
located at the ATCA position. The bright galaxy south of the ATCA
position has $z=0.0871$ from the 2dFGRS.

\noindent
{\bf NVSS~J230035$-$363410}: The very faint $K-$band identification
is offset by ($\Delta \alpha, \Delta \delta)$=(1\farcs4E, 1\farcs6S)
from the ATCA position.

\noindent
{\bf NVSS~J230123$-$364656}: The faint $K-$band identification is
located at the ATCA position.

\noindent
{\bf NVSS~J230203$-$340932}: The $K-$band identification is located
near the central radio component, which is most likely the core. In
table~2, we list the position of this radio core.

\noindent
{\bf NVSS~J230404$-$372450}: The $K-$band identification is located
between the radio lobes.

\noindent
{\bf NVSS~J230527$-$360534}: No $K-$band source is seen near the
radio position in our SofI image.

\noindent
{\bf NVSS~J230822$-$325027}: The $K-$band identification is located
at the ATCA position.

\noindent
{\bf NVSS~J230846$-$334810}: The radio source has a complex
morphology, consisting of at least 4 components. The $K-$band
identification is surrounded by a number of fainter clumps. This
source resembles WN~J1015+3038 and TN~J1049$-$1258 \citep{deb02}.

\noindent
{\bf NVSS~J230954$-$365653}: There are several faint $K-$band sources
located in between the radio lobes. A deeper radio map would be needed
to determine with certainty which one is the host galaxy. We
tentatively identify the galaxy closest to the line connecting the
radio lobes.

\noindent
{\bf NVSS~J231016$-$363624}: The bright $K-$band identification is
located between the radio lobes. This source is also detected in the
DENIS survey with $K=14.197$ in a 4\farcs5 aperture.

\noindent
{\bf NVSS~J231144$-$362215}: The very faint $K-$band identification
is located between the radio lobes.

\noindent
{\bf NVSS~J231229$-$371324}: The $K-$band identification is located
at the ATCA position.

\noindent
{\bf NVSS~J231311$-$361558}: The $K-$band identification is located
at the midpoint of the radio lobes. This is the largest radio source
in our sample.

\noindent
{\bf NVSS~J231317$-$352133}: The position quoted in
Table~\ref{datatable} is for the bright eastern component only,
because the $K-$band identification coincides with this lobe.

\noindent
{\bf NVSS~J231335$-$370609}: The $K-$band identification is offset
by ($\Delta \alpha, \Delta \delta)$=(0\farcs3E, 1\farcs9N) from the
ATCA position.

\noindent
{\bf NVSS~J231338$-$362708}: The faint $K-$band identification is
located between the radio lobes.

\noindent
{\bf NVSS~J231341$-$372504}: The $K-$band identification is located
at the midpoint of the radio lobes.

\noindent
{\bf NVSS~J231357$-$372413}: The $K-$band identification is located
at the ATCA position.

\noindent
{\bf NVSS~J231402$-$372925}: The $K-$band identification is located
at the ATCA position.

\noindent
{\bf NVSS~J231459$-$362859}: The $K-$band identification is located
at the ATCA position.

\noindent
{\bf NVSS~J231519$-$342710}: The $K-$band identification is located
along the radio axis, offset by ($\Delta \alpha, \Delta
\delta)$=(5\farcs9E, 2\farcs1S) from the midpoint between the radio
lobes in the ATCA image.

\noindent
{\bf NVSS~J231726$-$371443}: There are two $K-$band objects located
along the radio axis, which may both be components of the host
galaxy. We conservatively assume that only the northernmost component,
offset by ($\Delta \alpha, \Delta \delta)$=(0\farcs2E, 2\farcs2S) from
the northern radio lobe in the ATCA image, is the host galaxy. The position
quoted in Table~2 is for the bright northern component
only, because it is near the NVSS position.

\noindent
{\bf NVSS~J231727$-$352606}: No $K-$band source is seen near the
radio position in our medium deep SofI image.

\noindent
{\bf NVSS~J232001$-$363246}: The faint $K-$band identification is
located at the ATCA position.

\noindent
{\bf NVSS~J232014$-$375100}: The $K-$band identification is located
between the radio lobes of this large radio source, offset by ($\Delta
\alpha, \Delta \delta)$=(4\farcs1W, 0\farcs1S) from the NVSS position.

\noindent
{\bf NVSS~J232058$-$365157}: The $K-$band identification is located
at the ATCA position.

\noindent
{\bf NVSS~J232100$-$360223}: The very faint $K-$band identification
is offset by ($\Delta \alpha, \Delta \delta)$=(1\farcs2W, 1\farcs1N)
from the ATCA position.

\noindent
{\bf NVSS~J232219$-$355816}: No $K-$band source is seen near the
radio position in our SofI image.

\noindent
{\bf NVSS~J232322$-$345250}: The $K-$band identification is located
at the ATCA position.

\noindent
{\bf NVSS~J232408$-$353547}: An optical galaxy is seen in both the DSS
and 2MASS at the NVSS position, so we did not observe this source with
the ATCA or AAT.

\noindent
{\bf NVSS~J232602$-$350321}: The bright $K-$band identification is
coincident with an extended radio source.

\noindent
{\bf NVSS~J232651$-$370909}: The faint $K-$band identification is
located at the ATCA position.

\noindent
{\bf NVSS~J232956$-$374534}: The faint $K-$band identification is
offset by ($\Delta \alpha, \Delta \delta)$=(0\farcs7E, 1\farcs5N) from
the ATCA position.

\noindent
{\bf NVSS~J233558$-$362236}: We identify the host galaxy as the bright
$K-$band source, located along the radio axis, nearer to the brightest
radio lobe.

\noindent
{\bf NVSS~J233729$-$355529}: The radio source has a complex
morphology, consisting of three non-aligned components. We identify
the host galaxy with a diffuse $K-$band source, coincident with the
brightest radio component.

\noindent
{\bf NVSS~J234137$-$342230}: No $K-$band source is seen near the
radio position in our SofI image.

\noindent
{\bf NVSS~J234145$-$350624}: The bright $K-$band identification is
located at the ATCA position. This source is also known as
PKS~J2341$-$3506.

\noindent
{\bf NVSS~J234904$-$362451}: The $K-$band identification is located
between the radio lobes.

\noindent
{\bf NVSS~J235137$-$362632}: No $K-$band source is seen near the
radio position in our medium deep SofI image.

\end{document}